\documentclass[aps,prc,superscriptaddress,nofootinbib,twocolumn,showpacs]{revtex4-1}
\usepackage{amsmath}
\usepackage{graphicx}
\usepackage{amssymb}
\usepackage{multirow}
\usepackage{adjustbox}
\usepackage{hyperref}

\begin{document}

\title{Leveraging the Urysohn Lemma of Topology for an Enhanced Binary Classifier}

\author{E. L\'{o}pez Fune}
\email{elopez@nexialog.com}
\affiliation{Nexialog Consulting, \\ 110 Av. de la République, 75011 Paris, France.}

\begin{abstract}
In this article we offer a comprehensive analysis of the Urysohn's classifier in a binary classification context. It utilizes Urysohn's Lemma of Topology to construct separating functions, providing rigorous and adaptable solutions. Numerical experiments demonstrated exceptional performance, with scores ranging from $95\%$ to $100\%$. Notably, the Urysohn's classifier outperformed CatBoost and KNN in various scenarios. Despite sensitivity to the $p-$metric parameter, it proved robust and adaptable. The Urysohn's classifier's mathematical rigor and adaptability make it promising for binary classification, with applications in medical diagnosis, fraud detection and cyber security. Future research includes parameter optimization and combining the Urysohn's classifier with other techniques. It offers an elegant and principled approach to classification, ensuring integrity and valuable data insights.
\end{abstract}


\maketitle

\section{Introduction}
\label{Section_01}

Binary classification \cite{Fisher:1936fra} is a fundamental task in Machine Learning (hereafter ML), aiming to categorize instances into two discreet classes based on their features. Accurate classifiers are crucial in various domains, including healthcare, finance, cyber security and image recognition. Traditional approaches often rely on feature engineering, model selection, and optimization techniques to improve classification performance. However, there is an ongoing quest to explore novel methodologies that can further enhance classification accuracy and generalization.

In the ever-expanding landscape of data science, where the influx of information shows no signs of abating, the quest to comprehend complex data structures has become a paramount challenge. Central to this endeavor is the transformation of data instances into a geometric framework, where each data point finds its place in a multidimensional coordinate space. This perspective transcends the abstract realm of data instances, enabling us to harness the power of algebra and geometry to navigate through high-dimensional data with unprecedented clarity. However, due to the very abstract nature of the data, the challenge lies in not only comprehending the data itself but also in discovering the intrinsic order that underlies the apparent chaos. Herein lies the crucial role of Topology, a branch of mathematics that specializes in the study of shapes and structures, and provides a rich framework for understanding the properties of spaces and their transformations.

One fundamental theorem within Topology is the Urysohn's Lemma, which holds great potential for advancing binary classification algorithms. The lemma establishes a fundamental property of normal spaces, which are widely studied in Topology. By harnessing its power, it becomes possible to construct continuous functions that effectively separate different subsets within a given space. In this article, we propose a novel binary classifier that incorporates this lemma. Our classifier aims to exploit the intrinsic structural characteristics of the data by leveraging topological insights. By applying it, we seek to enhance the discriminatory capabilities of the classifier, leading to improved classification accuracy and robustness.

Existing binary classifiers often rely on explicit feature representations or assume certain underlying assumptions about the data distribution. While these approaches have achieved considerable success, they may struggle with complex, high-dimensional datasets or fail to capture subtle nonlinear relationships between features. By introducing the Urysohn's Lemma into the classification framework, we aim to address these limitations and provide a more flexible and expressive model that can better adapt to diverse datasets. This article presents the theoretical foundation of the lemma in the context of binary classification. We describe our proposed classifier, highlighting its integration into decision-making processes. Furthermore, we conduct comprehensive experiments on various benchmark datasets to evaluate the performance of our classifier against state-of-the-art methods. The results demonstrate the efficacy of leveraging this fundamental result from Topology in improving classification accuracy and robustness. By integrating this lemma into binary classification, we aim to advance the field by offering a new perspective and methodology for tackling challenging classification problems. Through empirical analysis and rigorous evaluation, we demonstrate the potential of this approach to unlock improved performance and expand the horizons of binary classification algorithms. All computational tasks were executed within a Python 3.7 environment, utilizing hardware resources that included an AMD Ryzen 7 3700u processor with Radeon Vega mobile graphics (8 cores), a RAM of 16 GB, and an AMD Radeon Vega 10 graphics card. The entire computational setup operated seamlessly on a 64-bit Ubuntu 22.04.3 LTS platform.

The remainder of this article is organized as follows. Section 2 provides an overview of the Urysohn's Lemma and its significance in Topology. Section 3 describes the methodology of our binary classifier, emphasizing the incorporation of the lemma. Section 4 includes the experimental setup and performance evaluation on benchmark datasets. Section 5 discusses the results and provides insights into the strengths and limitations of our proposed approach. Finally, Section 6 concludes the article by summarizing the key findings and discussing potential future directions for research.

\section{The Urysohn's Lemma of Topology}
\label{Section_02}
Topology, among other subjects, studies the properties of a geometric object that are preserved under continuous deformations, such as stretching, twisting, crumpling, and bending. Two fundamental concepts in Topology are closeness and normality. Understanding these concepts is essential for grasping the implications of the Urysohn's Lemma, which is the core of this section.

\subsection{Brief overview of normal and metric spaces in Topology}

A topological space $(X,\mathfrak{X})$ is structure consisting on a non-empty set $X$ endowed with a family of subsets of $X$: $\mathfrak{X}$, called a Topology, which allows defining continuous deformation of subspaces, and, more generally, all kinds of continuity. The elements of $\mathfrak{X}$ are called open sets, and satisfy three specific rules\footnote{There are several ways to define a Topolgy on a non-empty set $X$, but for the purposes of this article, we will stick with this one.}
\begin{enumerate}
\item The empty set and $X$ itself belong to $\mathfrak{X}$.
\item Any arbitrary (finite or infinite) union of members of $\mathfrak{X}$ belongs to $\mathfrak{X}$.
\item The intersection of any finite number of members of $\mathfrak{X}$ belongs to $\mathfrak{X}$.
\end{enumerate} 

A subset $C\subseteq X$ is said to be closed in $(X,\mathfrak{X})$ if its complement $X\setminus C \in \mathfrak{X}$ (is an open set). In the context of Topology, a closed set is like a group of points that includes its boundaries. This concept is of fundamental importance to what follows in this article, and used repeatedly in the application of Urysohn's lemma.

A \emph{Normal space} is a topological space that satisfies the normality property: any two disjoint closed subsets can be separated by disjoint open sets. This property ensures that points and sets can be separated within the space using open sets. Normal spaces are often used to establish separation axioms in Topology and are closely related to other properties such as compactness and Hausdorffness.

A metric space $(X,d)$ is a structure consisting on a non-empty set $X$ and a function $d: X\times X \mapsto \mathbb{R}$, called distance or metric, that satisfies four axioms for all points $x, y, z \in X$:
\begin{enumerate}
\item Reflexivity : $d(x,x)=0$.
\item Positivity: $d(x,y)>0 \Leftrightarrow x\neq y$.
\item Symmetry: $d(x,y)=d(y,x)$.
\item Triangle inequality: $d(x,z)\leq d(x,y)+d(y,z)$.
\end{enumerate}

A definition that we are going to use throughout this article is the definition of distance from a point to a set, and on which distance-based algorithms such as k-Nearest Neighbors (hereafter kNN) \cite{Cover:1967thp}(and references therein) and KMeans \cite{Lloyd:1982spm, Steinhaus:1957sth} are based. The distance from a point $x\in X$ to a subset $A\subset X$ is defined as follows:
\begin{align}
d(x,A) = inf\{d(x,a): a\in A\},\label{Eq_01}
\end{align}
\noindent where $inf({r})$ denotes the infimum (the greatest lower bound) of a numeric set. If the point $x$ belongs to $A$, then by its definition, $d(x,A)=0$.

In the literature there are different functions that have the properties of a metric, some of the best known are the so-called $p-$metrics in $\mathbb{R}^{n}$:
\begin{align}
d(x,y) = \left(\sum_{i=1}^{n}(x_i - y_i)^p\right)^{1/p},\label{Eq_02}
\end{align}
\noindent where $x_i, y_i$ are the $i-$th coordinates of the points $x, y \in \mathbb{R}^{n}$, and based on this metric, we will build our binary classifier.

\bigskip
Starting from a metric space, one can always construct a topological space based on the concept of open balls. An open ball of radius $r>0$ and centered at the point $x\in X$ is defined as:
\begin{align}
B_{r}(x) = \{y\in X: d(x,y) < r\}.\label{Eq_03}
\end{align}
\noindent It turns out, then, that for all points of $X$, the union of open balls centered on themselves behaves as if they were open in a Topology. We then have that every metric space induces a topological space, but the inverse is not always true, and there are many examples in the literature. 

\bigskip
Normal spaces provide a notion of separation and generalization of metric spaces, and it is an essential concept in Topology that lay the foundation for the Urysohn's Lemma. The lemma establishes a connection between normal spaces and the construction of continuous functions that effectively separate subsets within these spaces. On the other hand, metric spaces are endowed with a distance function, which measures the separation between points. This distance metric, along with the inherent Hausdorff ($T_2$) property \cite{Stephen:1970ste, Dugundji:1966top}, ensures that any two distinct points can be separated by disjoint open sets. Moreover, metric spaces adhere to the $T_4$ separation axiom \cite{Stephen:1970ste, Dugundji:1966top}, which goes beyond the basic Hausdorff property, ensuring that disjoint closed sets can be completely separated by open sets. Therefore, the combination of the metric and these separation properties inherently makes all metric spaces normal, making them a foundational and well-behaved class of topological spaces.

In the following sections, we will delve deeper into the theoretical implications of the Urysohn's Lemma and explore its application in the context of binary classification.

\subsection{Urysohn Lemma: Relevance and Implications in Topology}

The Urysohn Lemma is a fundamental theorem in Topology that plays a significant role in understanding the properties of spaces and their transformations. It furnishes us with an alternative way of characterizing the separation property of spaces, concerned with separating closed sets with open sets. The lemma states that:
\begin{quote}
A topological space $(X,\mathfrak{X})$ is normal, if and only if, for any two nonempty closed disjoint subsets $A,\,B \in \mathfrak{X}$, there is a continuous function $f: X\mapsto [0,1]$ such that $f(A)={0}$ and $f(B)={1}$. A function $f$ with this property is called a Urysohn function. 
\end{quote}
\noindent The demonstration can be found in many books of General Topology \cite{Stephen:1970ste, Dugundji:1966top}, but it is out of the scope of the present article.

The lemma demonstrates the existence of a continuous function that assigns distinct values to disjoint closed subsets. In other words, it shows that there is a continuous function that can effectively discriminate between different subsets within a normal space. This has profound implications in various areas of Topology and, as we shall see, in the context of binary classification as well.

Given a metric space $(X,d)$ which induces automatically a topology on $X$ with the normality property, one can construct a Urysohn's function, to separate two disjoint subsets $A, B \subset X$:
\begin{align}
f(x) = \dfrac{d(x,A)}{d(x,A) + d(x,B)}.\label{Eq_04}
\end{align}
\noindent It can be shown that this function satisfies $f(A) = 0$ and $f(B) = 1$ by using Eq.\eqref{Eq_01}, thus establishing the Urysohn's Lemma, which along with Eq.\eqref{Eq_04}, forms the theoretical backbone for the integration of topological insights into binary classification.

The function $f(x)$ defined in Eq.\eqref{Eq_04}, which assigns values in the interval $[0, 1]$ to points in the space $X$ based on their proximities to the two closed subsets $A$ and $B$, shares similarities with a probability measure and can be interpreted as such, but it's not necessarily a probability distribution in the formal sense. Indeed, $f(x)$ can be interpreted as a measure of how ``close'' or ``related'' the point $x$ is to subset $A$ compared to subset $B$. When $f(x)$ is closer to 1, it indicates that $x$ is more related to subset $B$, and when $f(x)$ is closer to 0, it suggests that $x$ is more related to subset $A$. This similarity to probability arises because the values are normalized to the range $[0, 1]$, just like probabilities. However, for this to be considered a probability distribution in the formal sense, it would need to satisfy additional properties, such as the total measure of $X$ being equal to 1. In other words, the sum or integral of $f(x)$ over the entire space $X$ should equal 1 for it to be a proper probability distribution. According to Eq.\eqref{Eq_04} and Eq.\eqref{Eq_02} for the $p-$metric used, it doesn't guarantee that the total measure will be 1 unless $X$ is bounded and non-empty. Therefore, while the function $f(x)$ has some similarities to probability measures in that it assigns values between 0 and 1 and can be interpreted as a measure of ``closeness'', it may not satisfy all the properties required for a formal probability distribution unless additional conditions are met. It depends on the specific properties of the subsets $A$, $B$, and the space $X$. However, it is a useful tool for defining proximity or relative closeness between points and subsets.

\section{Methodology and implementation}
\label{Section_03}

The Urysohn's Lemma, through the Urysohn's separating function, offers an innovative avenue for binary classification. By harnessing the power of Topology, it provides a fresh perspective on data separation, enriching the classification process with its ability to uncover concealed patterns and deliver superior classification outcomes.

\subsection{Description of the proposed binary classifier utilizing the Urysohn Lemma}

The Urysohn's Lemma, as presented in the previous section, is a powerful concept in Topology that finds an intriguing application in binary classification problems. In this context, it offers a unique approach for effectively separating data points into two distinct classes, making an analogy with the closed subsets $A$ and $B$. Central to this application is the introduction of a Urysohn's separating function $f(x)$, a continuous function designed to encapsulate the discriminatory characteristics of the data, and plays a pivotal role in classifying data points. Its primary objective is to assign each data point a real number that reflects its affinity with one of two classes. This assignment is based on a simple principle: data points belonging to one class receive a value equal to zero, while those of the other class are assigned a value equal to one. By smoothly transitioning between these two extremes, the Urysohn's separating function provides a continuous spectrum of values, allowing for a nuanced and adaptable classification process. It accomplishes this by leveraging the inherent topological structure of the data space, making it particularly valuable for complex and high-dimensional datasets.

In practice, the Urysohn's separating function aids in discerning patterns and relationships within data, even when they are subtle or intricate. This topological approach enhances classification accuracy and robustness, shedding light on hidden structures and anomalies that might otherwise remain obscured in the vast landscape of data. Data instances with a binary target variable can be effectively modeled as two closed and non-intersecting subsets in a finite-dimensional real metric space using a method that leverages topological concepts. Data instances can be represented as points in a finite-dimensional real metric space, where each point has a finite number of coordinates, making it suitable for working with real-world data that can be represented numerically. In binary classification problems, one has a target variable that can take one of two values, often denoted as 0 and 1. This target variable serves as the basis for splitting the data into two subsets that represent the two classes one wants to classify. Mathematically, one can identify these subsets with closed subsets within the finite-dimensional real metric space, as any finite collection of points contains all its limit points. To model the two classes as non-intersecting, one must ensure that there are no data points that belong to both subsets. This implies that the boundary between the two classes is well-defined, and each point is uniquely associated with one and only one of the binary classes. By representing binary classification data as two closed and non-intersecting subsets within a finite-dimensional real metric space, one can establish a mathematical foundation for the classification problem. This modeling approach allows for the application of the Urysohn's Lemma, to facilitate the separation of these two classes with the help of a Urysohn's separating function, enhancing the accuracy and robustness of binary classification. We put all this in practice in the next subsections.

\subsection{Overview of the classifier's architecture, algorithms, and techniques employed}

The UrysohnClassifier\footnote{Documentation Jupyter Notebook: \url{https://github.com/elopezfune/Urysohn-s-Binary-Classifier}} is a non-parametric binary classifier that leverages the principles of Topology, specifically Urysohn's Lemma, to separate and classify data points. This classifier is designed for applications where distinguishing between two classes is crucial and where the topological structure of the data plays a significant role in the separation process. Let's break down the architecture, algorithms, and techniques employed by the UrysohnClassifier:

\begin{enumerate}
\item Initialization: 
\begin{itemize}
\item The classifier's constructor, \textbf{\textunderscore\textunderscore init\textunderscore\textunderscore}, initializes with a choice of distance metric (e.g., Euclidean, Manhattan and $p-$metric), the control parameter $\epsilon$, and the number of permutations for a feature importance assessment.
\end{itemize}
\item Training:
\begin{itemize}
\item  The \textbf{fit} method is used to train the classifier on a labeled dataset. It stores the training data for later use.
\end{itemize}
\item Classification:
\begin{itemize}
\item \textbf{predict\textunderscore proba}: This method, first computes the distance (using a p-metric) between each data instance and each of the two labeled training subsamples. Then, for each data instance, it evaluates the Urysohn's separating function Eq.\eqref{Eq_04}, to predict the degree of closeness of a data point belonging to one of the two classes, providing a continuous affinity value for classification.
\item \textbf{predict}: The `predict` method classifies data points based on the closeness estimates. It assigns the data points to one of the two classes, taking the class with the higher affinity, according to a threshold, usually taken to a probability of $50\%$.
\end{itemize} 
\item Permutation-Based Feature Importance: 
\begin{itemize}
\item \textbf{permutation\textunderscore importance}: This method assesses the importance of each feature in the classification process. It does this by permuting the values of individual features and observing the impact on classification performance. The results are stored in a DataFrame, providing insights into feature importance.
\end{itemize}
\item Data Structures and Libraries: 
\begin{itemize}
\item The classifier utilizes data structures such as NumPy arrays and Pandas DataFrames to handle data efficiently.
\item It also relies on well-established libraries like NumPy, Pandas, and Scikit-Learn (e.g., `roc\textunderscore auc\textunderscore score`) to perform various calculations and assessments, and its architecture was created to be compatible with other functionalities of the latter library.
\end{itemize}
\end{enumerate}

In terms of time and space complexity, the UrysohnClassifier during the training phase the time complexity is $O(1)$ as it simply stores the training datapoints, and the space complexity would be $O(N*M)$ where $N$ represents number of datapoints and $M$ represents the number of features that determine each datapoint. During the predicting phase, the time complexity is $O(N)$, as for a single datapoint prediction, the computation of the distance to each training datapoint is required, and the minimum distance is retained. This represents a huge advantage over kNN, for example, where the k-nearest neighbors are required, being its time complexity $O(N*k)$. Regarding handling nominal data, the UrysohnClassifier is primarily designed for continuous data, and handling nominal data may require preprocessing (e.g., one-hot encoding or converting to a suitable metric space representation). 

On the other hand, the choice of metric may indirectly impact the classifier's performance. Since the UrysohnClassifier doesn't directly exhibit a classic bias-variance trade-off in the same way as many other ML algorithms, the bias in this classifier is primarily controlled by the choice of the Urysohn's separating function, which is determined by the selected distance metric. The separating function is essentially a continuous decision boundary between the two classes, and the bias introduced depends on the complexity of this boundary and how well it captures the underlying data patterns. In addition, the choice of distance metric also indirectly influences variance of the classifier, as some distance metrics may make the separating function more sensitive to noise or fluctuations in the data. Indeed, the choice of whether $p$ is even or odd in the $p-$metric can significantly impact how binary classification problems, including those addressed by the UrysohnClassifier, are affected by outliers, poor sampling, and class imbalance. 

In the case of $p$ even, the $p-$metric is sensitive to the squared values of the coordinates of the vector that represents the datapoint. This makes it more robust against extreme outliers and may provide a more stable separating function when dealing with unbalanced data. Alternatively, if $p$ is odd, the $p-$metric is more sensitive to the absolute values of the coordinates of the vector, making it more sensitive to outliers and less resistant to extreme values, which may be less robust and less effective in situations with imbalanced class distributions, as it can overemphasize the influence of a few dimensions, thus increasing the variance. An alternative solution would be to add a control parameter $+\epsilon$ to the denominator of the separating function $f(x)$ in Eq.\eqref{Eq_04}.

Experimentation and cross-validation are essential to determine which $p-$metric is most appropriate for a given problem, as the impact of $p$ can vary from one dataset to another. In the following section, a series of numerical experiments will be conducted to test the stability and robustness of the UrysohnClassifier for binary classifications.

\section{Experimental Setup}
\label{Section_04}

In this section, we will engage in conducting several numerical experiments with the binary classifier UrysohnClassifier to test its decision-making capabilities, study its stability, sensitivity, and robustness. We will utilize three well-known open-source datasets from the literature for conducting these experiments.

\subsection{Datasets}

The datasets used for the numerical experiments are classical open-source ML compiled tables for binary classification purposes. Three multivariate datasets were used: the Breast Cancer Wisconsin (Diagnostic) \cite{Wolberg:1995onw}, the Chronic Kidney Disease \cite{Rubini:2015lpp} and the Banknontes Annotation datasets \cite{Lohweg:2013vba}. In Table \ref{tab:Table_01}, under the acronyms {\bf BCW}, {\bf CKD} and {\bf BNA} respectively, are recorded some of their main characteristics.

\begin{table}[h!t]
\centering
\adjustbox{max width=\textwidth}{
\begin{tabular}{|c|c|c|c|}
\hline
\textbf{ }                       &     \textbf{BCW}       &       \textbf{CKD}              &        \textbf{BNA}               \\
\hline
\textbf{Samples}                 &         569            &            400                  &             1371                  \\
\textbf{Dimension}               &          30            &             25                  &                6                  \\
\textbf{Numerical variables}     &          28            &             11                  &                5                  \\
\textbf{Nominal variables}       &           2            &             14                  &                1                  \\
\textbf{Classes}                 &           2            &              2                  &                2                  \\
\textbf{Samples per class (0/1)} &     212/357            &        150/250                  &          761/610                  \\
\textbf{Imbalance ratio}         &       59.38$\%$        &             60$\%$              &           124.75$\%$              \\
\textbf{Missing values}          &          No            &            Yes                  &               No                  \\
\hline
\end{tabular}
}
\caption{Datasets main characteristics.}
\label{tab:Table_01}
\end{table}

The {\bf BCW} dataset is a well-known binary classification dataset based on features extracted from digitized images of breast mass fine needle aspirates (FNA). These features describe cell nuclei characteristics in the images. The dataset employs a separating plane obtained through the Multisurface Method-Tree (MSM-T), a classification method utilizing linear programming to construct decision trees. Relevant features are selected through an exhaustive search, considering 1-4 features and 1-3 separating planes. The {\bf CKD} dataset, donated on 7/2/2015, contains a range of medical tests and diagnostic procedures be performed on patients, and collected from hospitals nearly 2 months of period. This dataset was previously pre-processed by the author\footnote{\url{https://github.com/elopezfune/Chronic_Kidney_Disease}}, including data formatting, asymmetry corrections, imputation of missing values, one-hot-encoding of nominal variables, standardization and normalization of numerical variables, resulting in an increase in its dimensionality from 25 to 44. Regarding the {\bf BNA} dataset, it was derived from authentic and counterfeit banknote-like specimens, which were subjected to image acquisition. For digitization purposes, an industrial-grade camera commonly employed for print inspection was utilized, resulting in final images of $400\times 400$ pixels. The imaging process, facilitated by specific object lens configurations and object-to-camera distances, yielded grayscale images with an approximate resolution of 660 dots per inch (dpi). To extract features from these images, the Wavelet Transform tool was employed, and the variance, skewness, kurtosis and the entropy of the images were collected.

\subsection{Robustness Testing}

For the datasets introduced in the preceding subsection, we produced three models UrysohnClassifier (one for each dataset), with an $L_1$ or Manhattan metric ($p=1$), and control parameter $\epsilon=0$. This will be the parameter settings of our model unless explicitly stated. Prior to model creation, we partitioned each dataset into separate training and test subsamples, adhering to the conventional $70\%$ training and $30\%$ testing split commonly employed in ML applications. Subsequently, we trained the three models using the training subsamples. To assess their performance, we evaluated several scoring metrics for each model, including the accuracy score and ROC-AUC, utilizing the respective test subsamples. It is essential to acknowledge that, given the intrinsic characteristics of the separation function in Eq.\eqref{Eq_04}, the computation of any scoring metric using the training subsample will consistently yield the most favorable expected outcome (e.g. ROC-AUC$\sim1$), unless the $p-$metric used says otherwise. As a result, the assessment of the classifier's predictive and discriminative capabilities is primarily contingent on the test subsample. Therefore, underfitting is not a concern with this classifier.

\begin{figure}[h!t]
\centering
\includegraphics[width=0.45\textwidth]{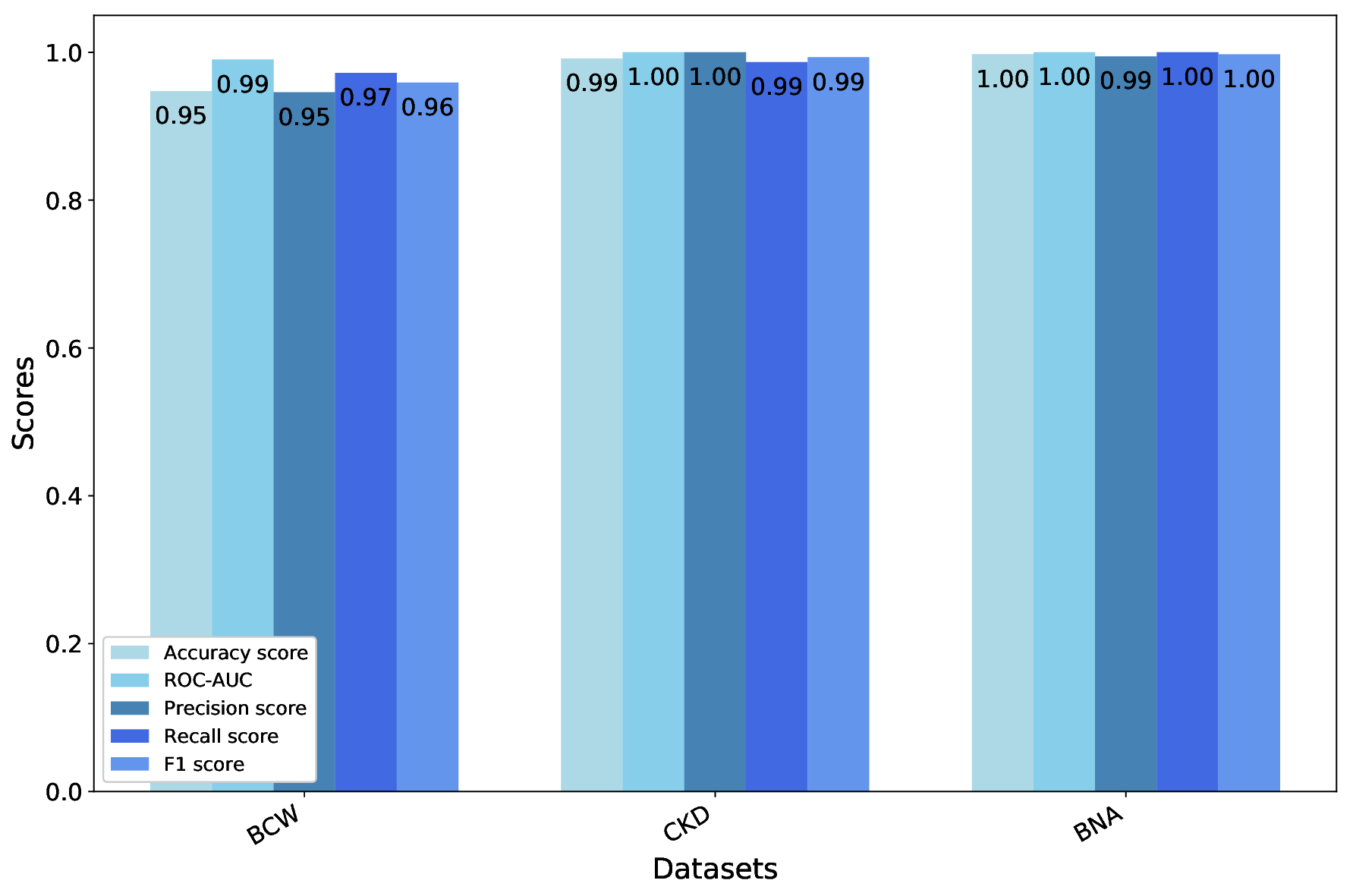}\\
\caption{(Color online) Score metrics performance of the binary classifier on the three datasets \textbf{BCW}, \textbf{CKD}, and \textbf{BNA}.} \label{Fig_01}
\end{figure}

The Fig.\eqref{Fig_01} illustrates the performance of each model associated with its respective dataset, as measured by various established scoring metrics. It is important to note that the scoring metrics were exclusively computed using the test subsample, as elucidated in the preceding paragraph. This approach was adopted because all scoring metrics yielded a constant value of 1 when calculated with the training subsample, for the chosen value of $p$. 

The evaluation of ML models is a critical aspect of model development, and the comprehensive use of multiple scoring metrics plays a pivotal role in assessing their performance. By thoroughly evaluating these models with a battery of scoring metrics, including accuracy score, ROC-AUC, precision, recall, and F1 score, we gained a multifaceted understanding of their predictive capabilities. The results obtained from these metrics for the three models provide valuable insights into their quality and predictive power. The first model exhibited a high level of accuracy ($95\%$) and strong performance across all metrics, with an F1 score of $96\%$, indicating a well-rounded model. The second model surpassed the first one, achieving near-perfect scores in accuracy and ROC-AUC ($99\%$ and $100\%$, respectively), emphasizing its exceptional predictive capabilities. The third model displayed exemplary performance, with perfect accuracy and ROC-AUC scores ($100\%$) and an F1 score of $100\%$, signifying its outstanding predictive power. The results obtained demonstrate the high-quality and robust predictive capabilities of the UrysohnClassifier, especially the second and third models, which exhibit remarkable performance across all metrics. This classifier holds promising potential for applications in real-world classification tasks, underscoring their utility in diverse domains. In addition, the three models do not present visible signs of overfitting. 

\begin{figure}[h!t]
\begin{center}
\includegraphics[width=0.35\textwidth]{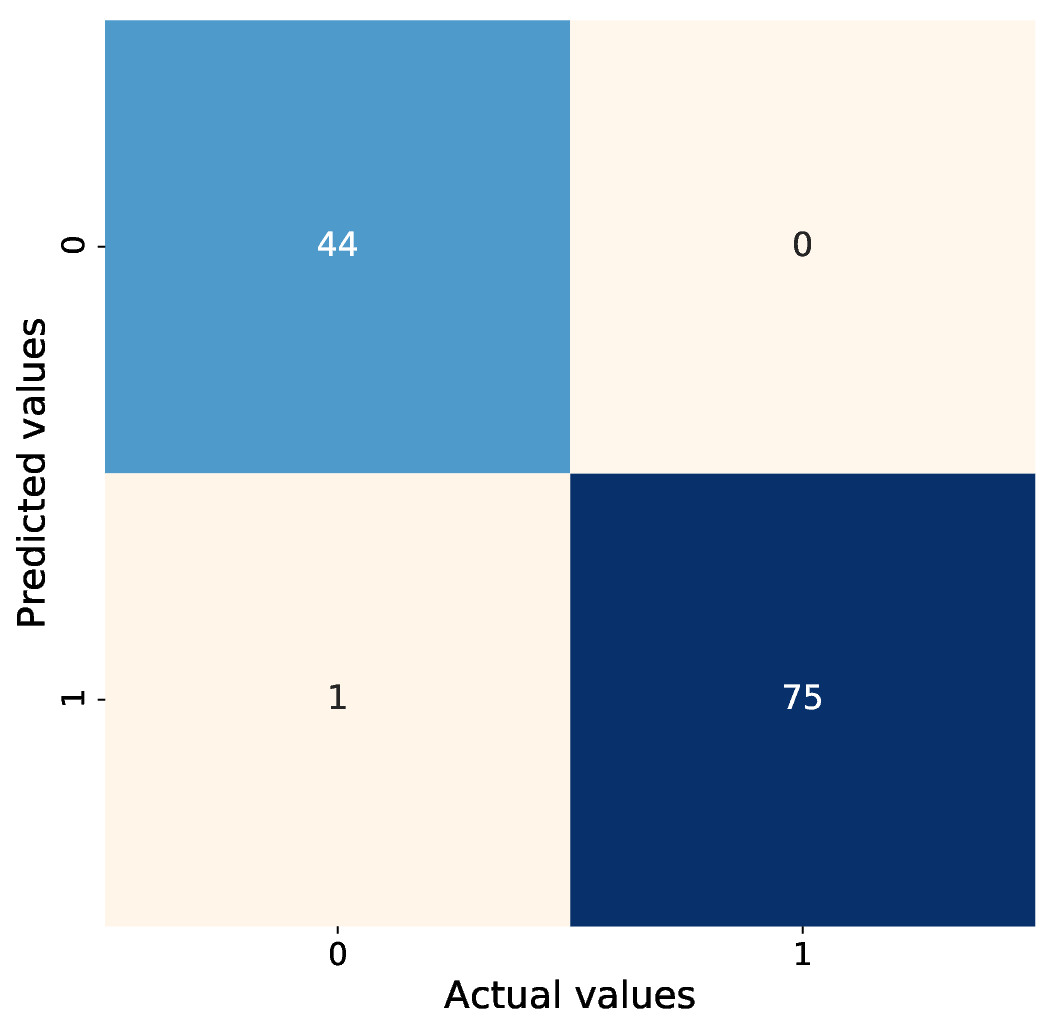}\\
\includegraphics[width=0.35\textwidth]{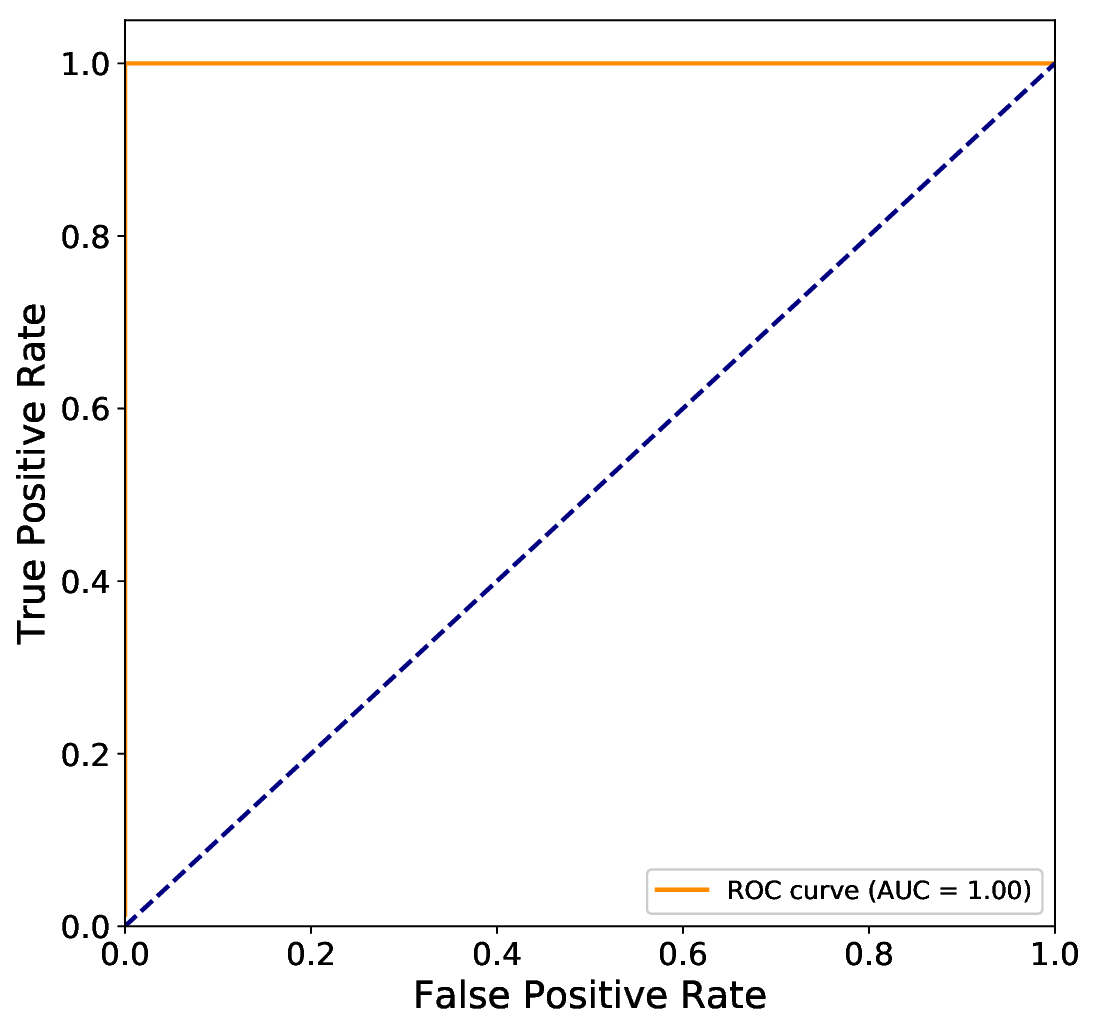}
\end{center}
\caption{(Color online) Confusion matrix (upper panel), ROC and AUC (lower panel) of the test subsamples.}\label{FIG_0203}
\end{figure}

\bigskip
For the following numerical experiments, we consider only the {\bf CKD} dataset since from the modeling point of view, there is a better compromise between performance and dimensionality. A new model is produced with the same parameters $p=1$ and $\epsilon=0$, and in this case, the confusion matrix and the curve ROC are shown in Fig.\eqref{FIG_0203}.

\begin{figure}[h!t]
\begin{center}
\includegraphics[width=0.5\textwidth]{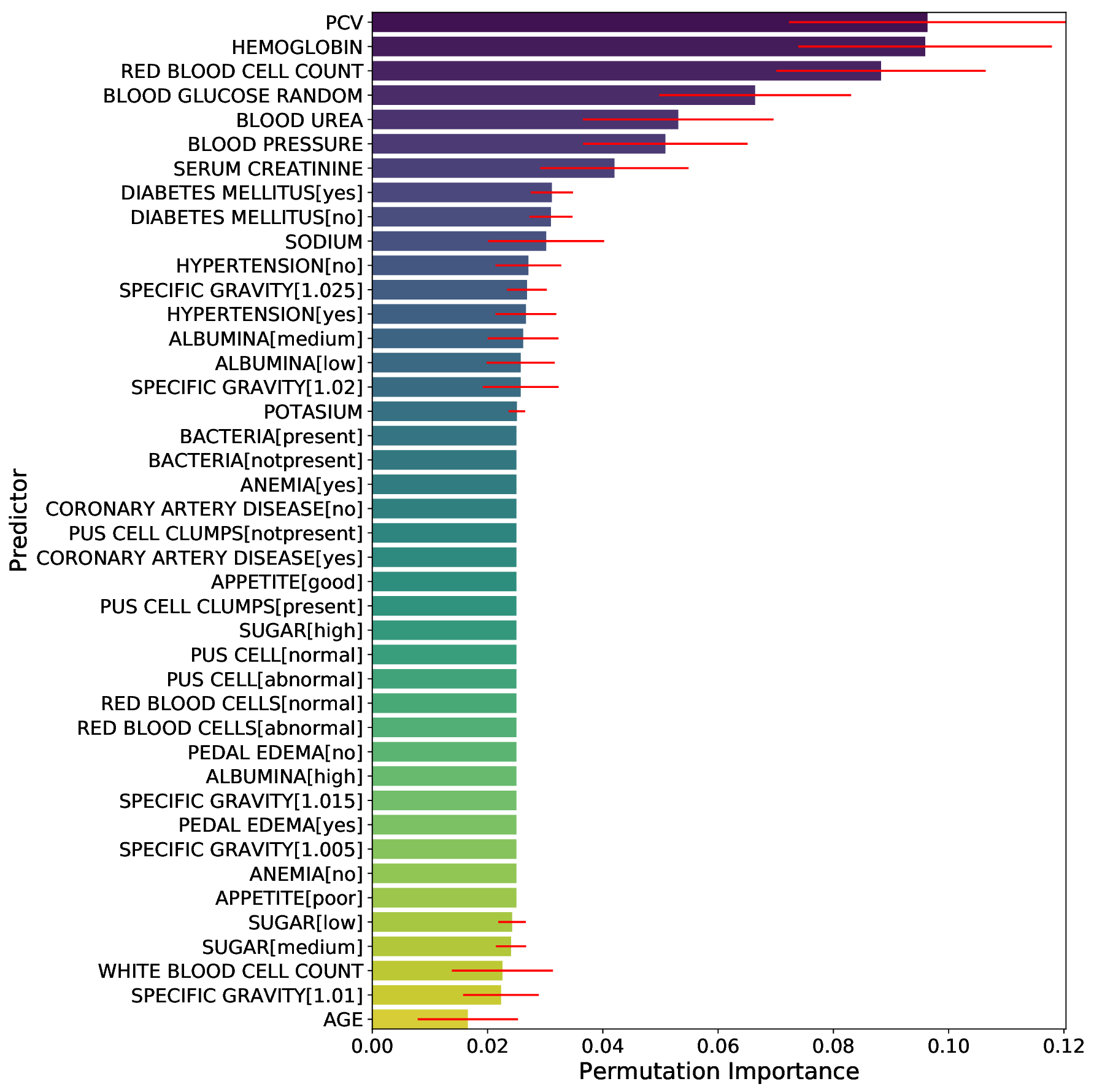}\\
\end{center}
\caption{(Color online) Variable importance using the permutations.}\label{FIG_04}
\end{figure}

The comparative analysis of the confusion matrices for this model and an optimized and cross-validated CatBoost model\footnote{Read the documentation Jupyter Notebook in \url{https://github.com/elopezfune/Chronic_Kidney_Disease}}\cite{Dorogush:2017age} on the same dataset reveals an intriguing observation. While both models achieved remarkably similar overall performances, with no differences in the number of false positives or negatives, it is essential to acknowledge the inherent complexity contrast between the two. The UrysohnClassifier offers an elegant and straightforward approach to binary classification, emphasizing simplicity and interpretability, as shown as well in Fig.\eqref{FIG_04}. In contrast, the CatBoost model, despite achieving a slightly better result in terms of false positives, is inherently much more intricate, involving a sophisticated ensemble of decision trees and complex optimization techniques. This juxtaposition underscores a compelling trade-off between model complexity and performance. The UrysohnClassifier, despite its comparative simplicity, showcases competitive results, underlining its efficiency and effectiveness in practical classification tasks. This highlights the significance of selecting a model that aligns with the specific requirements of a given task, considering the trade-offs between performance gains and the complexity inherent in the model's design and implementation.

The remarkable similarity in permutation importance results implies that both models, the UrysohnClassifier and CatBoost, rely on similar sets of features to make accurate predictions. This suggests that, despite their inherent differences in complexity and design, both models prioritize comparable factors when distinguishing between classes in the dataset.

\subsection{Sensitivity tests}

\begin{figure}[h!t]
\begin{center}
\includegraphics[width=0.5\textwidth]{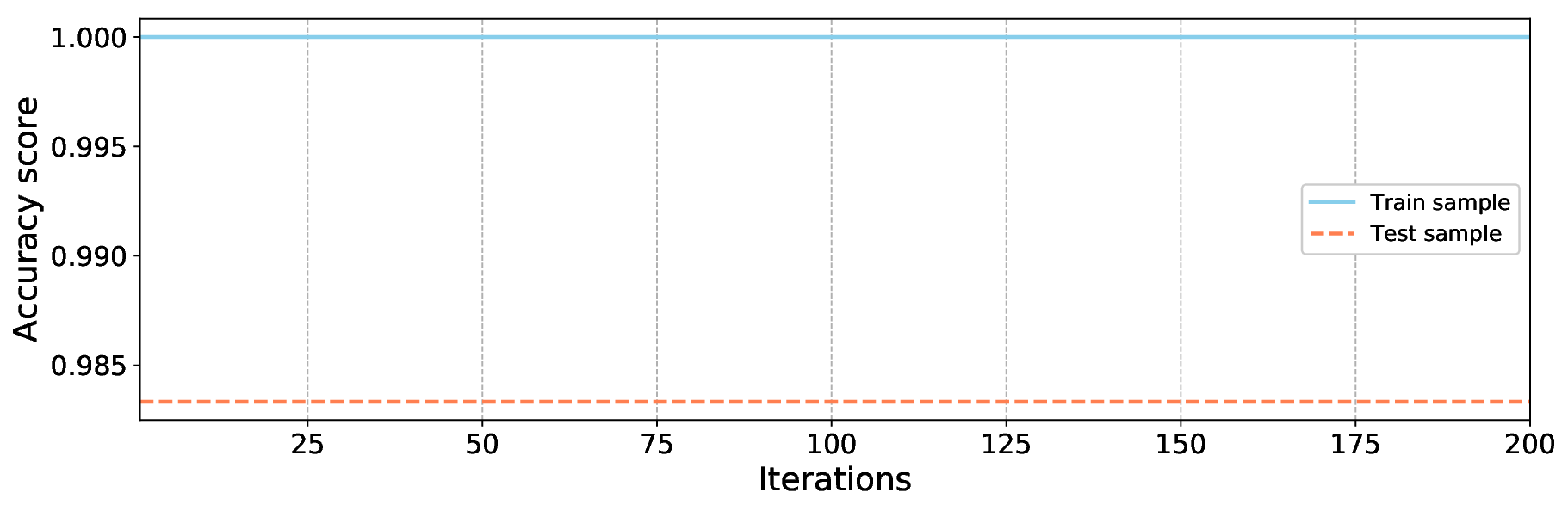}\\
\includegraphics[width=0.5\textwidth]{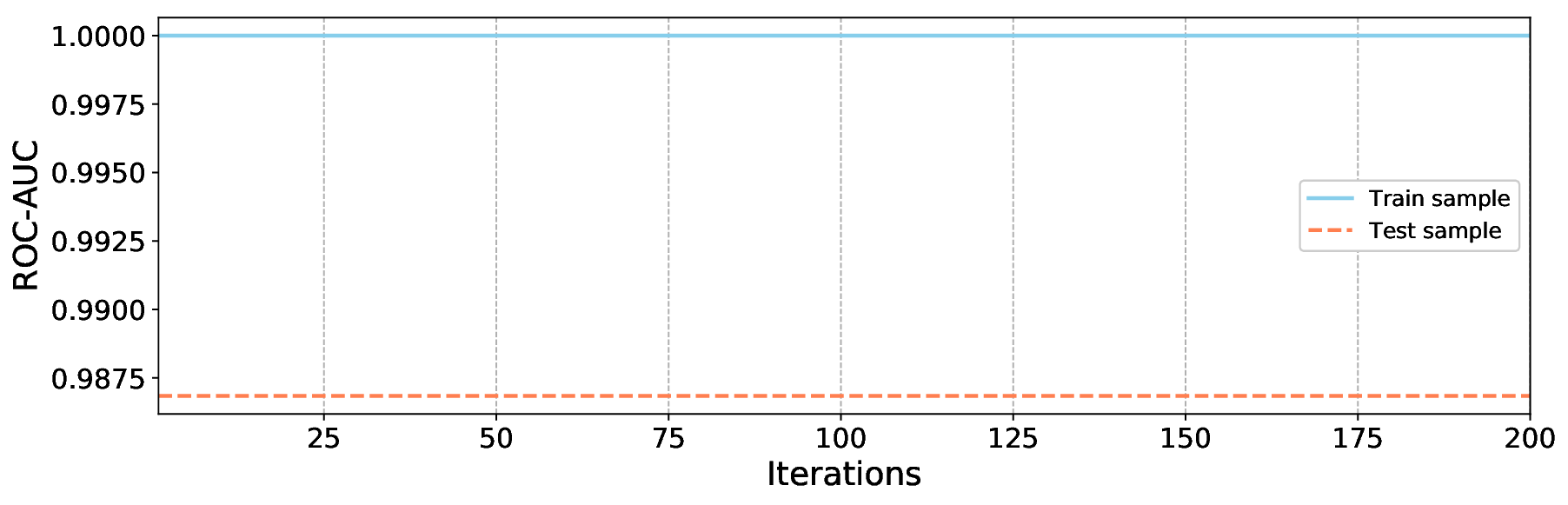}
\end{center}
\caption{(Color online) Sensitivity to randomness in train/test subsamples, measured by the accuracy score (upper panel) and the ROC-AUC (lower panel).}\label{FIG_0506}
\end{figure}

In the context of nonlinear problems, such as the one addressed in this article, it is crucial to acknowledge the potential influence of varying noise levels on the final outcomes. To assess the robustness and stability of the UrysohnClassifier, we conducted a series of iterative tests. In each iteration, we randomly selected training and testing subsamples, enabling us to create a new $p=1,\,\epsilon=0$ model and assess its performance based on scoring metrics, specifically Accuracy score and the ROC-AUC. The ensuing results, presented in Fig.\eqref{FIG_0506}, offer valuable insights into the classifier's consistency in the face of continuous alterations in training and test subsamples, underscoring its resilience in handling varying noise levels.

In a similar vein, we conducted another iterative experiment, this time focusing on the parameter $p$ of the chosen metric. We systematically varied $p$ within the range (1, 9), and for each distinct $p$ value while keeping $\epsilon=0$, we proceeded to create new training and testing subsamples. This approach allowed us to simultaneously perform cross-validation and parameter tuning, introducing an element of randomness during each subsampling iteration to assess the classifier's robustness against potential selection biases. Once again, we assessed the performance of each newly generated model using the Accuracy score and ROC-AUC based on the $p$ parameter value.

\begin{figure}[h!t]
\begin{center}
\includegraphics[width=0.5\textwidth]{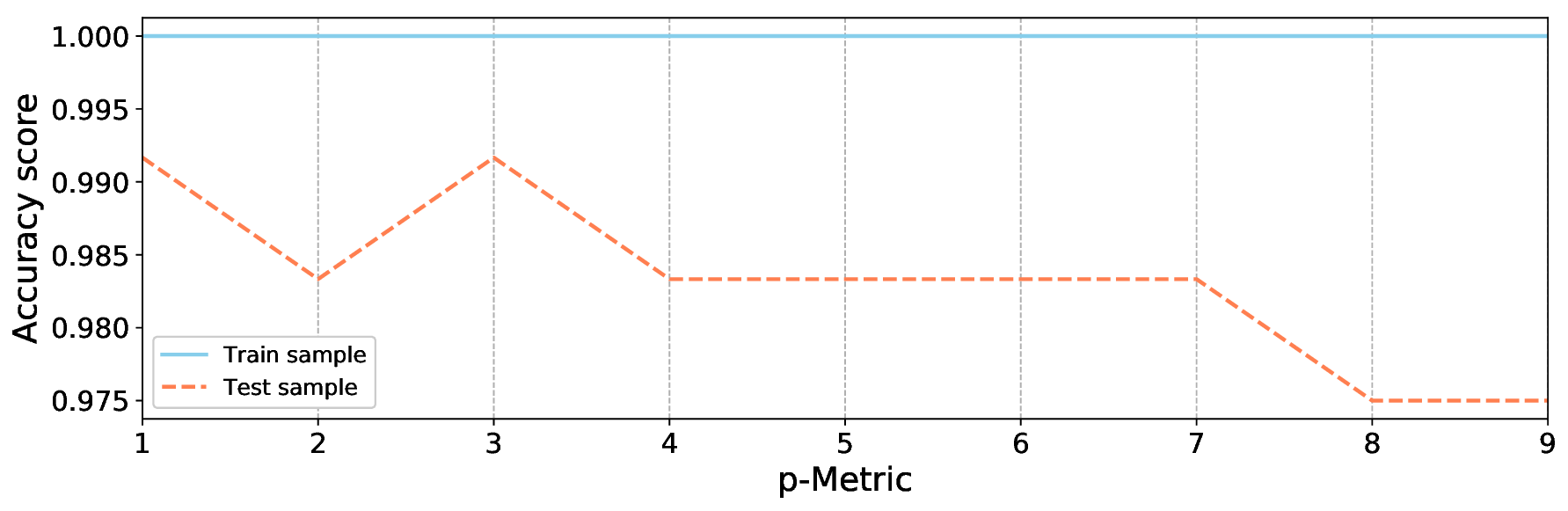}\\
\includegraphics[width=0.5\textwidth]{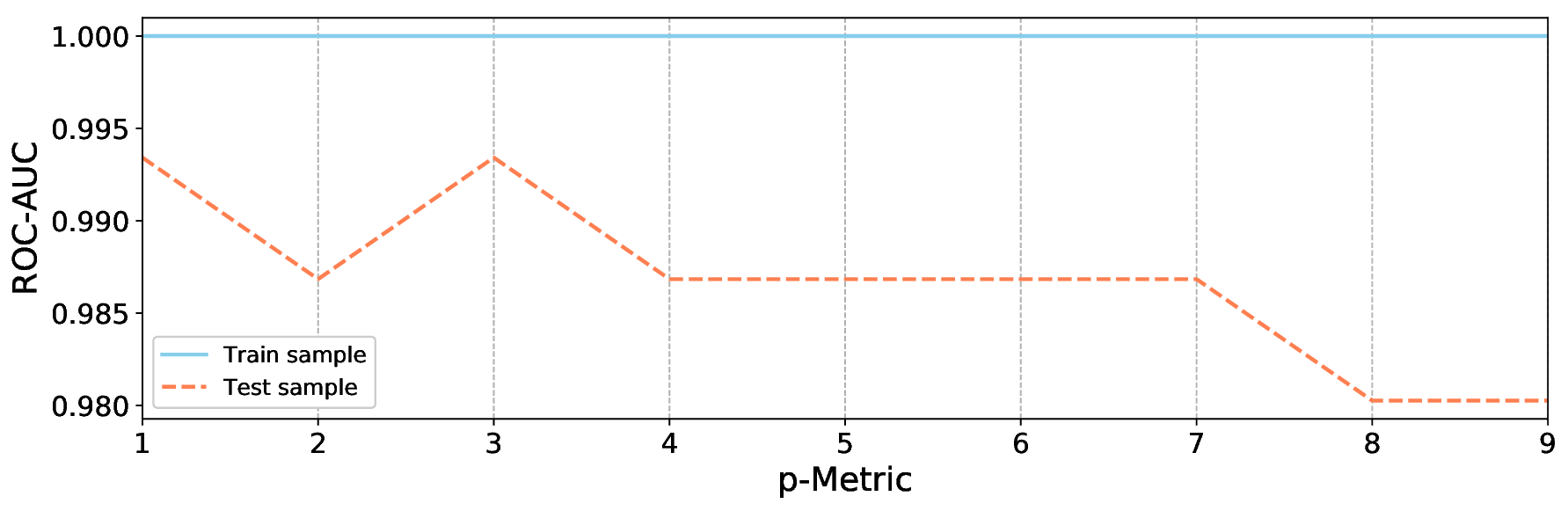}
\end{center}
\caption{(Color online) Sensitivity to the metric type, measured by the accuracy score (upper panel) and ROC-AUC (lower panel).}\label{FIG_0708}
\end{figure}

The results obtained from this experiment are particularly noteworthy, as they unveil a compelling trend. Models with low values of $p$ exhibit enhanced robustness, stability, and a reduced risk of overfitting when compared to models large values of $p$, at least within the context of this dataset. Notably, both scoring metrics, for the test subsample, consistently attain their maximum values at $p=1$ and $p=3$, while decreasing when $p$ keeps increasing. The most pronounced performance disparities between the training and test subsamples are observed for values of $p$ larger than $3$, indicative of potential overfitting, and we believe this is due to the sensitiveness to outliers and extreme values, rather than to the classifier itself. Importantly, our experiments did not reveal any indications of underfitting throughout the analysis. These findings shed light on the influence of the $p$ parameter on model performance and underscore the importance of selecting an appropriate value of $p$ to optimize classifier performance.

\subsection{Comparative Analysis}

\begin{figure}[h!t]
\begin{center}
\includegraphics[width=0.5\textwidth]{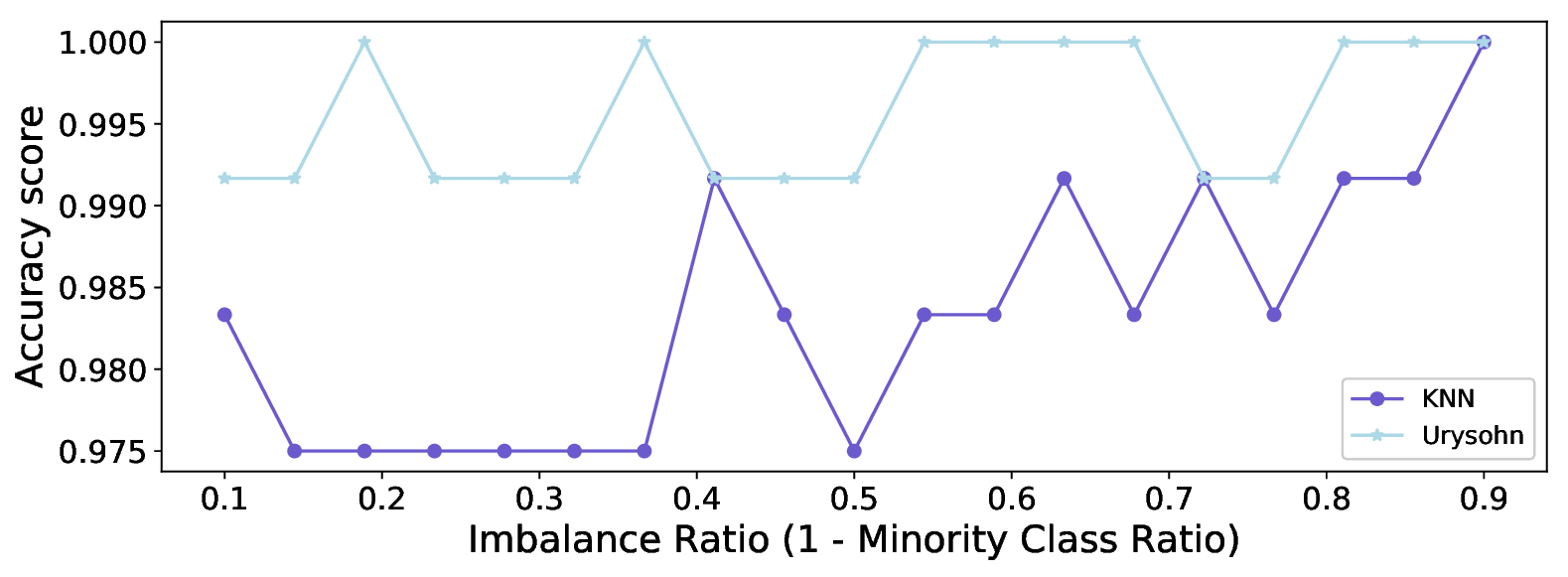}\\
\includegraphics[width=0.5\textwidth]{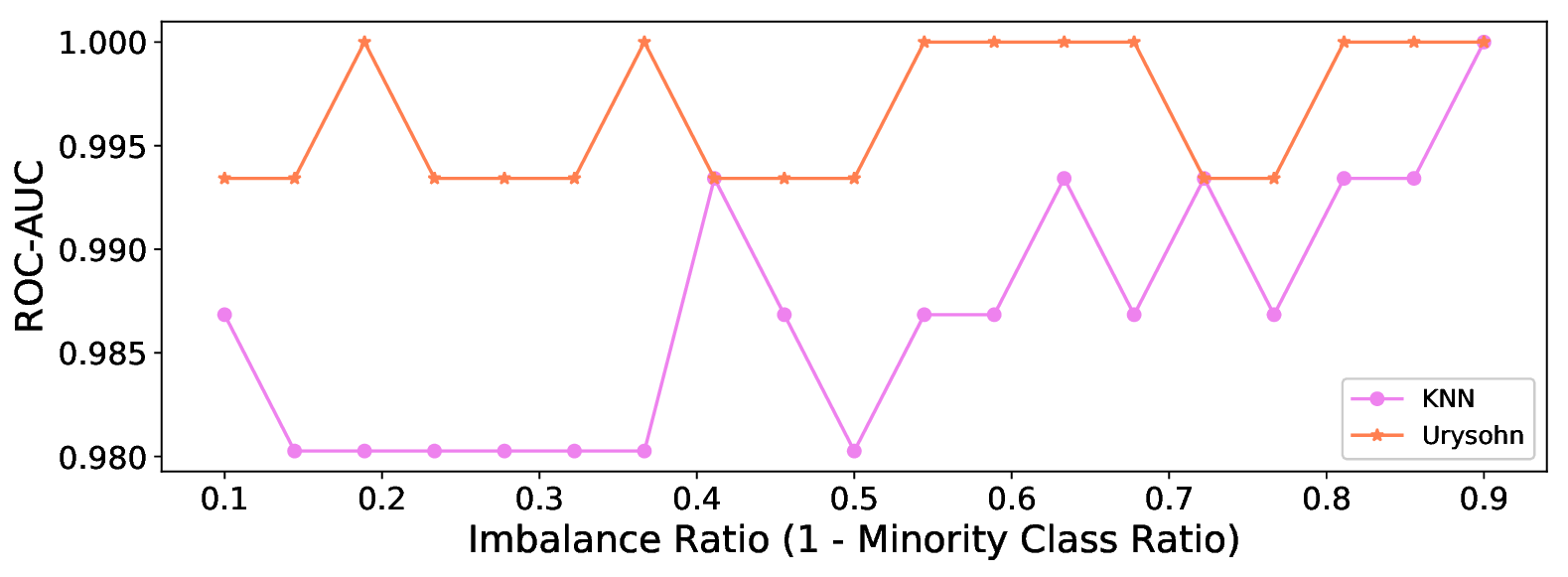}
\end{center}
\caption{(Color online) Sensitivity to class imbalance, measured by the accuracy score (upper panel) and ROC-AUC (lower panel).}\label{FIG_0910}
\end{figure}

In this subsection, we conduct two sensitivity tests and a comparative analysis involving the UrysohnClassifier and a counterpart algorithm from the same family, namely the kNN. In the first test, we systematically assess the performance, as measured by the Accuracy score and the ROC-AUC, of both our classifier and the kNN ($k=3$) in scenarios characterized by an imbalance in the target variable within the training data. These evaluations are performed iteratively, with the imbalance ratio varying from $10\%$ to $90\%$ in increments of $10\%$. To address this imbalance, we apply resampling techniques to the training subsample, ensuring that the class 0 and class 1 imbalance aligns with the desired ratio for each iteration. Subsequently, we create new models with hyperparameters $p=1$ and $\epsilon=0$ for each imbalance ratio and compute the accuracy score and ROC-AUC using the respective test subsample. The resultant findings are visually presented in Fig.\eqref{FIG_0910}. 

Across both scoring metrics and every algorithm under consideration, a consistent pattern emerges, indicating enhanced performance as the degree of imbalance in the binary variable to be predicted diminishes. This trend aligns with the anticipated outcome and underscores the influence of class distribution on predictive accuracy. Of paramount significance is the discernible disparity in performance between our classifier and the kNN algorithm, particularly in scenarios characterized by a substantial class imbalance. This notable discrepancy underscores the substantial potential of our classifier within its algorithmic category, suggesting its efficacy in aiding decision-making processes, particularly when faced with highly imbalanced datasets.

\begin{figure}[h!t]
\begin{center}
\includegraphics[width=0.5\textwidth]{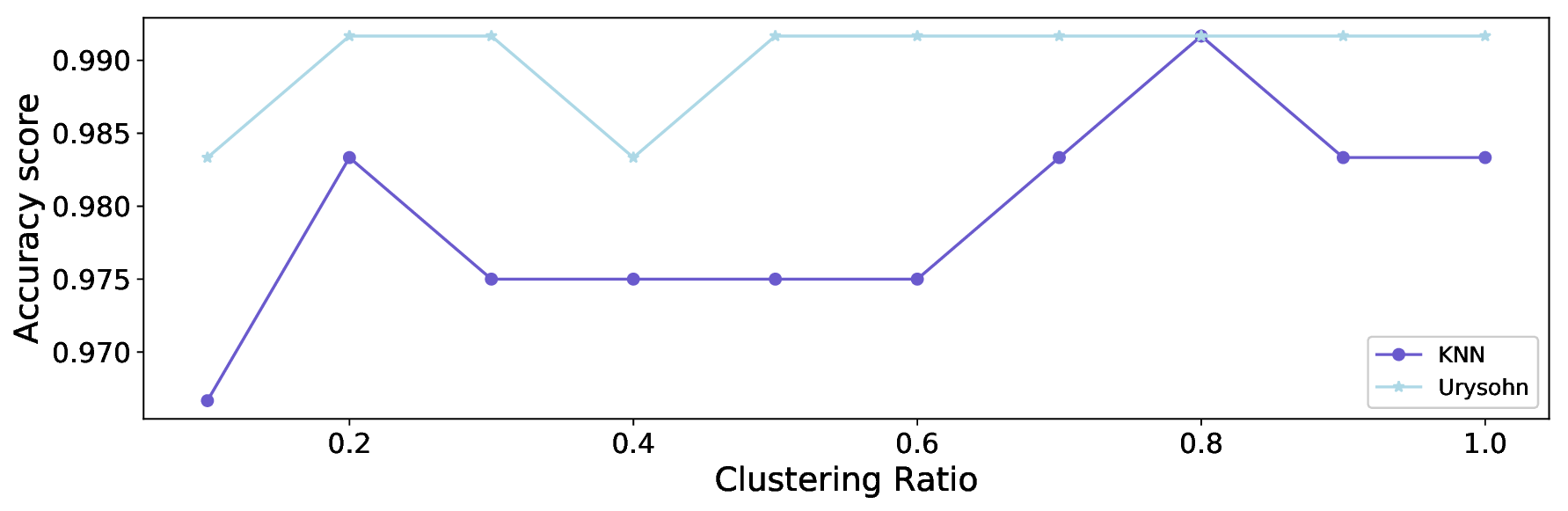}\\
\includegraphics[width=0.5\textwidth]{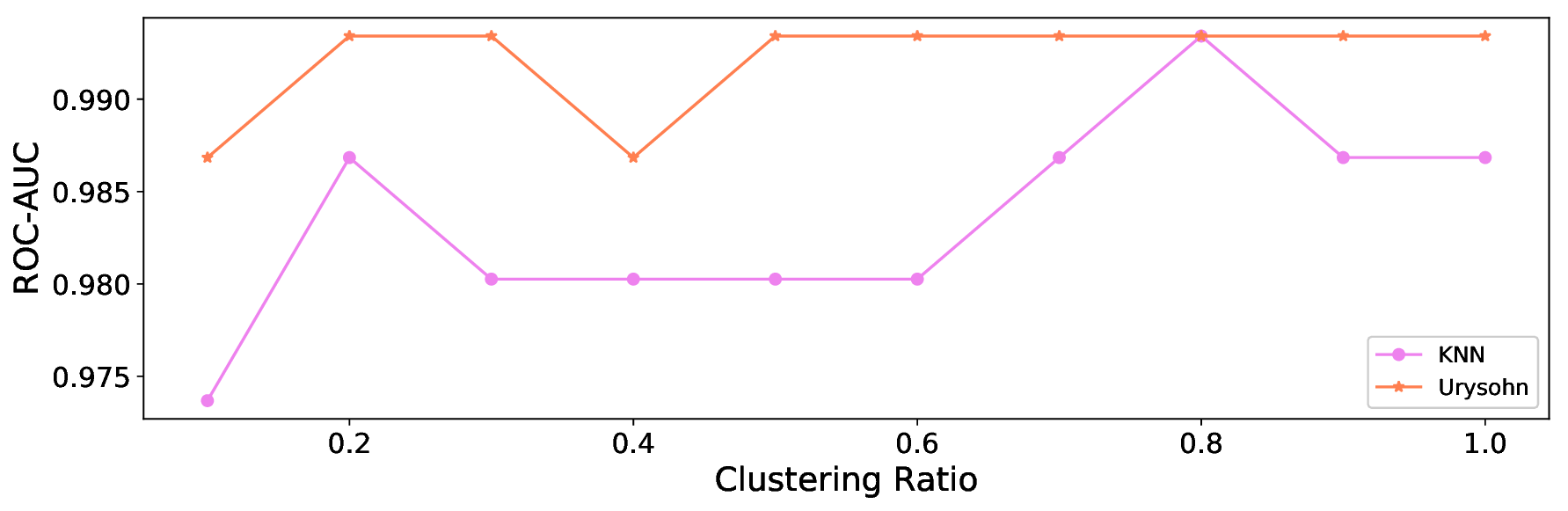}
\end{center}
\caption{(Color online) Sensitivity to data clustering, measured by the accuracy score (upper panel) and ROC-AUC (lower panel).}\label{FIG_1112}
\end{figure}

\bigskip
The second sensitivity test and a comparative analysis involves the data clustering. In a manner analogous to the previous scenario, we conducted this test in an iterative fashion. Initially, we partitioned the dataset into two subsamples: one designated for training the algorithm and model development (comprising $70\%$ of the total dataset), and the other reserved for assessing the scoring metrics (constituting $30\%$ of the total dataset). Subsequently, we assembled a list of clustering ratios ranging from $10\%$ to $100\%$, with increments of $10\%$. In each iteration, we introduced random resampling of the training subsample, adjusting its size in accordance with the clustering ratio. Throughout these iterations, we created fresh models with hyperparameters set to $p=1$ and $\epsilon=0$, and systematically assessed their performance using accuracy score and ROC-AUC metrics. This experimental approach aimed to gauge the sensitivity of our classifier to varying levels of data clustering and also allowed us to make a comparative evaluation against the kNN ($k=3$) algorithm. The outcomes of this test are presented in Fig.\eqref{FIG_1112}, shedding light on our classifier's performance under different clustering conditions.  

\section{Discussion}
\label{Section_05}

In this section, we delve into a comprehensive analysis of the results obtained from a series of numerical experiments involving the Urysohn's classifier within the context of binary classification. We discuss the strengths and weaknesses of this classifier, compare its performance with the CatBoost and kNN algorithms, address any encountered limitations, summarize key findings, explore the implications of leveraging Urysohn's Lemma for binary classification, and suggest potential avenues for future research.

\subsection{Performance}
The application of Urysohn's Lemma to construct a separating function for binary classification within the context of closed and non-intersecting subsets in a finite-dimensional real metric space can be well-justified for several compelling reasons. Urysohn's Lemma is a fundamental result in Topology, specifically designed for the purpose of separating disjoint closed sets within a topological space. By applying this well-established topological concept, we bring a robust mathematical foundation to binary classification, ensuring that the separation process is based on solid principles. This topological elegance provides a solid and principled approach to classification.

In our experiments with three diverse datasets, we observed exceptional results with metric scores ranging from $95\%$ to $100\%$ in terms of Accuracy, ROC-AUC, Precision, Recall, and F1 scores. Notably, the Urysohn's classifier exhibited its best performance with the {\bf BNA} dataset, followed by the {\bf CKD} dataset and lastly the {\bf BCW} dataset. Such results highlight the effectiveness of this classifier in various application scenarios. Moreover, in a second experiment, we compared the Urysohn's classifier with an optimized and cross-validated CatBoost model on the same dataset, and both showed equivalent performances with no differences between false positives and negatives. This is a noteworthy achievement, considering the algorithmic complexity of CatBoost. Furthermore, the Urysohn's classifier and CatBoost yielded similar feature importances, indicating the classifier's potential for accurate interpretation. 

In terms of robustness, a third experiment demonstrated that the Urysohn's classifier remains stable even in the presence of noisy data, as evidenced by consistently high scores against the iteration index. One of the most significant comparisons made in this study is between the Urysohn's classifier and the kNN algorithm. In both cases of class imbalance and varying clustering sizes, the Urysohn's classifier consistently outperformed kNN. This suggests that Urysohn's classifier, with its topological elegance, offers advantages over kNN in terms of performance. Additionally, the interpretability of the Urysohn's classifier is a noteworthy advantage. Unlike kNN, which relies on proximity and a potentially complex decision boundary, the Urysohn's classifier provides a clear and principled approach to separation. This can be invaluable in applications where interpretability is crucial, such as in finance, medical diagnosis, etc.

\subsection{Limitations, Findings, and Implications}

While the Urysohn's classifier exhibits several strengths, it is essential to acknowledge its limitations. The choice of the $p-$metric parameter can impact its performance, as our experiments showed that low and high values of $p$ result in different metric scores. This sensitivity to the $p-$metric parameter should be considered when applying it to different scenarios.

The key findings from our experiments indicate that the Urysohn's classifier is a robust, principled, and stable method for binary classification. It delivers strong performance, particularly when compared to kNN, and offers interpretability in complex datasets. Moreover, it is resilient to noise and class imbalance, making it a promising candidate for various real-world applications.

The application of Urysohn's Lemma to binary classification offers a novel and rigorous approach. By leveraging topological principles, it provides a mathematical foundation that can enhance the reliability and interpretability of classification models. This approach could find applications in various domains, including medical diagnosis, fraud detection, and many others where robustness and interpretability are essential.

\section{Conclusion}
\label{Section_06}

The application of Urysohn's Lemma to construct separating functions for binary classification has demonstrated its potential as a principled and adaptable approach. The discussion in the preceding section highlights the topological elegance and mathematical rigor that Urysohn's Lemma brings to the classification process. The crucial ability of the Urysohn's separating function to ensure non-intersecting classes is a fundamental aspect that prevents ambiguity and misclassification in binary classification tasks. It offers a clear distinction between different classes, improving the reliability of the classification outcomes. The continuous and smooth nature of the separating function allows for a nuanced and flexible approach to classification. It can capture complex relationships and subtle patterns within the data, providing valuable insights into the data structure.

Adaptability is another key strength of the Urysohn's classifier. It can adjust to the specific characteristics of the data without relying on fixed decision boundaries. This adaptability makes it a versatile choice for a wide range of datasets and real-world applications. In high-dimensional spaces, the topological foundation of Urysohn's Lemma becomes increasingly valuable. It can effectively address the challenges posed by complex, multidimensional data, making it a relevant choice for modern classification tasks.

Our study opens up exciting possibilities for future research. First, further investigation into the sensitivity of the Urysohn's classifier to the $p-$metric parameter could lead to strategies for parameter optimization. The generalization to multiclass classification and regression is as well an ongoing subject. Additionally, exploring the combination of the Urysohn's classifier with other ML techniques or ensembles may provide further performance improvements. Moreover, the application of Urysohn's Lemma in more complex and high-dimensional datasets remains an area ripe for exploration. Finally, the development of practical guidelines for applying the Urysohn's classifier in real-world applications is an important direction for future research.

In conclusion, Urysohn's Lemma offers an elegant, principled, and adaptable approach to binary classification. Its topological foundation ensures the integrity of the classification process, while its adaptability and continuous separation make it a valuable tool for data-driven decision-making.

\vspace*{5mm}


\end{document}